\documentclass[lettrpaper]{amsart}
\pdfoutput=1

\usepackage{amsfonts}
\usepackage{amsmath}
\usepackage{amssymb}
\newcommand{\be}{\begin{equation}}
\newcommand{\ee}{\end{equation}}
\newcommand{\dz}{\wedge}
\newcommand{\der}{{\rm d}}
\newcommand{\bbR}{\mathbb{R}}

\begin{document}
\title{How the green light was given for gravitational wave search}
\vskip 1.truecm \author{C Denson Hill} \address{Department of Mathematics, Stonybrook University, New York, USA}
\email{\tt dhill@math.stonybrook.edu}
\author{Pawe\l~ Nurowski} \address{Center for Theoretical Physics Polish Academy of Sciences, Warszawa, Poland}
\email{nurowski@fuw.edu.pl} \thanks{This research was supported by the
Polish National Science Centre under the grant DEC-2013/09/B/ST1/01799.}
\date{\today}
\begin{abstract}
The recent detection of gravitational waves by the LIGO/VIRGO team \cite{ligo} is an incredibly impressive achievement of experimental physics. It is also a tremendous success of the theory of General Relativity. It confirms the existence of black holes; shows that binary black holes exist; that they may collide and that during the merging process gravitational waves are produced. These are all predictions of General Relativity theory in its fully nonlinear regime.

The existence of gravitational waves was predicted by Albert Einstein in 1916 within the framework of linearized Einstein theory. Contrary to common belief, even the very \emph{definition} of a gravitational wave in the fully nonlinear Einstein theory was provided only after Einstein's death. Actually, Einstein had arguments against the existence of nonlinear gravitational waves (they were erroneous but he did not accept this), which virtually stopped development of the subject until the mid 1950s. This is what we refer to as the  \emph{Red Light} for gravitational waves research. 

In the following years,  the theme was picked up again and studied vigorously  by various experts, mainly Herman Bondi, Felix Pirani,  Ivor Robinson and Andrzej Trautman, where the  theoretical obstacles concerning gravitational wave existence were successfully overcome, thus giving the \emph{Green Light} for experimentalists to start designing detectors, culminating in the recent LIGO/VIRGO discovery. 

  In this note we tell the story of this theoretical breakthrough. Particular attention is given to the fundamental 1958 papers of Trautman \cite{T1,T2}, which seem to be lesser known outside the circle of General Relativity experts.  A more detailed technical description of these 2 papers is given in the Appendix.  
\end{abstract}
  \maketitle
\vspace{0.5cm}
\noindent
\section{GRAVITATIONAL WAVES IN EINSTEIN'S LINEARIZED THEORY}
    The idea of a gravitational wave comes directly from {\bf Albert Einstein} ({\bf 14.3. 1879-18.04.1955}). Immediately after formulating General Relativity Theory \cite{Einstein1916a}, still in 1916 \cite{Einstein1916b}, he linearized his field equations
    \be R_{\mu\nu}-\tfrac12 R g_{\mu\nu}=\kappa T_{\mu\nu}\label{eeq}\ee
    by assuming that the metric $g_{\mu\nu}$ representing the gravitational field has the form of a slightly perturbed Minkowski metric $\eta_{\mu\nu}$,
    $$g_{\mu\nu}=\eta_{\mu\nu}+\epsilon h_{\mu\nu}.$$
    Here $0<\epsilon\ll 1$, and the term `he linearized' means that he developed the left hand side of (\ref{eeq}) in powers of $\epsilon$ and neglected all terms involving $\epsilon^k$, with $k>1$. As a result of this linearization Einstein found the field equations of \emph{linearized} General Relativity, which can conveniently be written for an unknown
    $$ \bar{h}{}_{\mu\nu}=h_{\mu\nu}-\frac{1}{2}\eta_{\mu\nu}h_{\alpha\beta}\eta^{\alpha\beta}$$
    as
$$\square \bar{h}{}_{\mu\nu}=2\kappa T_{\mu \nu},\quad\quad \square=\eta_{\mu\nu}\partial^\mu\partial^\nu.$$   
%\centerline{\includegraphics[height=6cm,width=6cm]{Figures/ein_1916_1.jpg}}
    These equations, outside the sources where $$T_{\mu\nu}=0,$$ constitute a system of decoupled relativistic wave equations
\be \square h_{\mu\nu}=0\label{lee}\ee
    for each component of $h_{\mu\nu}$. This enabled Einstein to conclude that \emph{linearized} General Relativity Theory admits solutions in which the perturbations of Minkowski spacetime $h_{\mu\nu}$ are plane waves traveling with the speed of light. Because of the \emph{linearity}, by superposing plane wave solutions with different propagation vecors $k_\mu$, one can get waves having any desirable wave front. Einstein named these \emph{gravitational waves}. He also showed that within the linearized theory these waves carry energy, and found a formula for the energy loss  in terms of the third time derivative of the quadrupole moment of the sources.

    Since far from the sources the gravitational field is very weak, solutions from the linearized theory should coincide with solutions from the full theory. Actually the wave detected by the LIGO/VIRGO team was so weak that it was treated as if it were a gravitational plane wave from the linearized theory. We also mention that essentially all visualizations of gravitational waves presented during popular lectures or in the news are obtained using linearized theory only.\\
    
\section{THE RED LIGHT}
We focus here on the fundamental problem posed by Einstein in 1916, which bothered him to the end of his life. The problem is: Do the fully nonlinear Einstein equations admit solutions which can be interpreted as gravitational waves?

If yes, then far from the sources, it is entirely reasonable to use linearized theory. If no, then it makes no sense to expend time, effort and money to try to detect such waves: solutions from the linearized theory are not physical; they are artifacts of the linearization.

If the answer is `no' we refer to it as a `red light' for gravitational wave search. This red light can be switched to `green' only if the following subproblems are solved:
\begin{itemize}
\item[(1)] What is a definition of a \emph{plane} gravitational wave in the full theory?
\item[(2)] Does the so defined plane wave exist as a solution to the full Einstein system?
\item[(3)] Do such waves carry energy?
\item[(4)] What is a definition of a gravitational wave with \emph{nonplanar front} in the full theory?
\item[(5)] What is the energy of such waves ?
\item[(6)] Do there exist solutions to the full Einstein system satisfying this definition?
\item[(7)] Does the full theory admit solutions corrresponding to the gravitational waves emitted by bounded sources ?
\end{itemize}  
To give a green light here, one needs a satisfactory answer to all these subproblems. Let us explain: Suppose that only the questions (1)-(3) had been settled in a satisfactory manner. Can we have a `green light'? The answer is `no', because contrary to the linear theory, unless we are very lucky, there is no way of superposing plane waves to obtain waves with arbitrary fronts. Thus the existence of a plane wave does not mean the existence of waves that can be produced by bounded sources, such as for example binary black hole systems. 

\section{SEARCH FOR PLANE WAVES IN THE FULL THEORY}

\subsection{Naive approach}
    A naive answer to our question (1) could be: a gravitational plane wave is a spacetime described by a metric, which in some cordinates $(t,x,y,z)$, with $t$ being timelike, has metric functions depending on $u=t-x$ only; preferably these functions should be $\sin$ or $\cos$. That this is not a good approach is seen in the following example:

Consider the metric 
$$\begin{aligned} g=&(\eta_{\mu\nu}+h_{\mu\nu})\der x^\mu\der x^\nu=\der t^2 - \der x^2 - \der y^2 - \der z^2+
      \\& \cos(t - x)\Big(2 +\cos(t - x)\Big)\der t^2 - 
 2 \cos(t - x) \Big(1+\cos(t - x)\Big)\der t \der x +\\& 
 \cos^2(t - x) \der x^2. \end{aligned}$$
We see here that the terms in the second ant third rows give the perturbation $h_{\mu\nu}\der x^\mu\der x^\nu$ of the Minkowski metric $\eta=\eta_{\mu\nu}\der x^\mu \der x^\nu=\der t^2 - \der x^2 - \der y^2 - \der z^2$. They are \emph{oscillatory}, and one sees that the \emph{ripples of the perturbation move with speed of light}, $c=1$, along the $x$ axis. A closer look shows also that the coefficients $h_{\mu\nu}$ of the perturbation satisfy the wave equation (\ref{lee}) (since they depend on a single null coordinate $u$, only), and more importantly, that the full metric $g$ is \emph{Ricci flat}.

Thus the above metric is not only an example of a `gravitational wave' in the linearized Einstein theory, but also it provides an example of a solution of the vacuum Einstein equations $R_{\mu\nu}=0$ in the \emph{fully nonlinear} Einstein theory. With all this information in mind, in particular having in mind the sinusoidal change of the metric with the speed of light in the $x$ direction, we ask: is this an example of a plane gravitational wave?

The answer is \emph{no}, as we created the metric $g$ from the flat Minkowski metric $\eta=\der \bar{t}^2 - \der x^2 - \der y^2 - \der z^2$ by a \emph{change of the time coordinate}: $\bar{t}=t+\sin(t-x)$. In view of this, the metric $g$ is just the flat Minkowski metric, written in nonstandard coordinates. As such it does not correspond to any gravitational wave!

The moral from this example is that attaching the name of a `gravitational wave' to a spacetime which was only inspected in some coordinate system, in which the metric coefficients satisfied conditions whispered by our intuition, is a wrong approach. As we see in this example we can always introduce a sinusoidal behaviour of the metric cooeficients, and their `movement' with speed of light, by an appropriate change of coordinates.

More generally, we need some nonintuitive, mathematically precise definition of even a plane wave.

\subsection{Red light switched on: Einstein and Rosen}
The first ever attempt to define a plane gravitational wave in the full theory is due to Albert Einstein and {\bf Nathan Rosen} ({\bf 22.3.1909-18.12.1995}). It happened in 1937 \cite{ER}, twenty years after the formulation of the concept of a plane wave in the linearized theory. They thought that they had found a solution of the vacuum Einstein's equations in the full theory representing a plane polarized gravitational wave. They observed that their solution had certain singularities, and as such must be considered as \emph{unphysical}. Their opinion is explicitly expressed in the subsequent paper of Rosen \cite{R}, which has the following abstract: \emph{The system of equations is set up for the gravitational and electromagnetic fields in the general theory of relativity, corresponding to plane polarized waves. It is found that all nontrivial solutions of these equations contain singularities, so that one must conclude that strictly plane polarized waves of finite amplitude, in contrast to cylindrical waves, cannot exist in the general theory of relativity.} The Einstein-Rosen paper \cite{ER} was refereed by Howard P. Robertson, who recognized that the singularities encountered by Einstein and Rosen are merely due to a wrong choice of coordinates, and that, if one uses correct coordinate patches, the solution may be interpreted as a cylindrical wave, which is nonsingular everywhere except on the symmetry axis corresponding to an infinite line source. This is echoed in Rosen's abstract \cite{R} quoted above in his phrase `in contrast to cylindrical waves', and is also mentioned in the abstract of the earlier Einstein-Rosen paper \cite{ER}, whose first sentence is: \emph{The rigorous solution for cylindrical gravitational waves is given}. Nevertheless, despite the clue given to them by Robertson, starting from 1937, both Einstein and Rosen did not believe that physically acceptable plane gravitational waves were admitted by the full Einstein theory.  This belief of Einstein affected the views of his collaborators, such as Leopold Infeld, and more generally many other relativists. If a plane gravitational wave is not admitted by the theory, and if this statement comes from, and is fully supported by, the authority of Einstein, it was hard to believe at any fundamental level that the predictions of the linearized theory are valid.

\subsection{Towards the green light: Bondi, Pirani and Robinson}
It is now fashionable to say that a new era of research on gravitational waves started at the International Conference on Gravitation held at Chapel Hill on 
18-23 January 1957. To show that not everybody was sure about the existence of gravitational waves during this conference we quote the following \cite{B}: \emph{Polarized plane gravitational waves were first discovered by N. Rosen, who, however, came to the conclusion that such waves could not exist because the metric would have to contain certain physical singularities. More recent work by Taub and McVittie showed that there were no unpolarized plane waves, and this result has tended to confirm the view that true plane gravitational waves do not exist in empty space in general relativity. Partly owing to this, Scheidegger and I have both expressed the opinion that there might be no energy-carrying gravitational waves at all in the theory.} In this quote the subject `I' is {\bf Herman Bondi} ({\bf 1.11.1919-10.9.2005}), one of the founding fathers of gravitational wave theory, and he refers to his opinion expressed during the Chapel Hill Conference in January 1957. Interestingly, the quote is from Bondi's \emph{Nature} paper announcing the discovery of a singularity-free solution of a plane gravitational wave, received by the journal only \emph{two months later}, on 24th of March 1957.

Bondi in the \emph{Nature} paper invokes the solution of Einstein's equations found in the context of gravitational waves by {\bf Ivor Robinson} ({\bf 7.10.1923-27.05.2016}). That paper, and the subsequent paper written by Bondi, {\bf Felix Pirani} ({\bf 2.2.1928-31.12.2015}) and Robinson \cite{BPR}, answers in positive our problems (1), (2), and (3).

In particular (1) is answered with the following definition of a \emph{plane wave in the full theory}: The gravitational plane wave is a spacetime which (a) satisfies vacuum Einstein's equations $R_{\mu\nu}=0$ and which (b) has a 5-dimensional group of isometries. The motivation for this definition is the fact that a plane electromagnetic wave has a 5-dimensional group of symmetries. Bondi, Pirani and Robinson do \emph{not} assume that the 5-dimensional group of isometries is isomorphic to the symmetry of a plane electromagnetic wave. They inspect all Ricci flat metrics with symmetries of dimension greater or equal 4 given by Petrov \cite{Pet}, and find exactly one class of solutions with the same 5-dimensional group of isometries, which by a miracle \emph{is} isomorphic to the symmetry group of the electromagnetic field.
%The 5-dimensional group of isometries acts transitively on the spacetime, so the spacetime is a homogeneous space, which means that it is nonsingular at each point.

It follows that the class of metrics obeying the Bondi-Pirani-Robinson definition of a plane gravitational wave depends on two \emph{free functions of one variable} that can be interpreted as the wave amplitude and the direction of polarization. Using these free functions Bondi, Pirani and Robinson obtained a \emph{sandwich wave}, i.e. a gravitational wave which differs from the Minkowski spacetime only in a 4-dimensional strip moving in a given direction with the speed of light. They used this sandwich wave and analyzed what happens when it hits a system of test particles. It follows that the wave \emph{affects} their motion, which leads to the conclusion that \emph{gravitational plane waves in the full theory carry energy}.     

In this way, the \emph{Nature} paper of Bondi \cite{B}, together with the later paper of Bondi, Pirani and Robinson \cite{BPR}, \emph{solves our problems} (1), (2) \emph{and} (3): the plane wave in the full theory is defined, it is realized as a class of solutions of Einstein fields equations $R_{\mu\nu}=0$, and it carries energy, since passing through the spacetime in a form of a sandwich it affects test particles.

As a last comment in this section we mention that the Bondi-Pirani-Robinson gravitational plane waves, looked for with great effort by physicists for 40 years, were actually discovered already in 1924 by a \emph{mathematician} {\bf H. W. Brinkman} \cite{Br}. He discovered the so called \emph{pp-waves}, a class of Ricci flat metrics having radiative properties, which include Bondi-Pirani-Robinson plane waves as a special case. His discovery was published in English in \emph{Mathematische Annalen}. If only there had been better communication between mathematicians and physicists. 

\section{GENERAL GRAVITATIONAL WAVES}

\subsection{Closer to the green: Pirani}
    The development of the theory of gravitational waves at the turn of the 1950's and 1960's was very rapid. The story, as we are presenting it here now, is more topical then chronological, so breaking the chronology, we will now discuss an important paper of Felix Pirani \cite{Pi}, which appeared before Bondi's \emph {Nature} announcement of the existence of a plane wave in Einstein's theory. It is also worthwhile to note that Pirani's paper \cite{Pi} was submitted a few months \emph{before} the Chapel Hill conference. For us, this paper is of fundamental importance, since among other things, it gives the first attempt of a purely geometric \emph{definition of a gravitational wave spacetime}.

    Pirani in \cite{Pi} argues that gravitational radiation should be detectable by analysis of the Riemann tensor. He suggests that a spacetime in which gravitational radiation is present has its Weyl tensor (the Lorentz group invariant part of the Riemann curvature tensor, other than the Ricci tensor and its scalar) \emph{algebraically special}. This statement is based on the so called \emph{Petrov classification} of gravitational fields. A \emph{generic} 4-dimensional Lorentzian metric $g$ distinguishes, at every point of the spacetime, \emph{four null directions}, which are called \emph{principal null directions} (PNDs). They are defined at each point by the Weyl tensor of $g$. There are a number of \emph{degenerate} situations, starting with the Weyl tensor being identically equal to zero, when no principal direction is defined. In less degenerate cases some of the 4 principal null directions can coincide, defining pointwise the types of spacetimes. In particular, if two out of the four PNDs coincide, and two other are different from these two and from each other - the [2,1,1] case, the spacetime at this point is type II. We may also have the [3,1] case - type III, the [4] case - type $N$, and the [2,2] case - type $D$. The generic case is [1,1,1,1]. If the Weyl tensor is not generic, i.e., if it is one of the cases [2,1,1], [2,2], [3,1], [4], at a point (or in a region) of spacetime, the gravitational field at this point (or in this region), is called \emph{algebraically special}. The main suggestion of Pirani in \cite{Pi}, based on his analysis of the 4-velocity of an observer following a gravitational field, is that in the presence of radiation, in the \emph{wave zone}, such an observer will see that the gravitational field is \emph{algebraically special}. At that time Pirani did not have all the Petrov types spelled out correctly (the fully correct Petrov classification is due to {\bf Roger Penrose} \cite{Pen}), so he does not say if in the wave zone the gravitational field is of type $II$, $III$, $D$ or $N$. However, his intuition about the importance of algebraic speciality in the theory of gravitational waves was correct, and now it is known that far away from the sources the Weyl tensor of a radiative spacetime must be of type $N$.\\

\subsection{Switching on green: Radiation is nonlocal}
Pirani's algebraic speciality condition for a gravitational wave spacetime refers to pointwise defined objects - the PNDs. As the Weyl tensor can change its algebraic type from point to point, the criterion is local. On the other hand, even in Maxwell theory, radiation is a nonlocal phenomenon. To illustrate this we recall a well known conundrum:\\
    \emph{Q: Does a unit charge hanging on a thread attached to the ceiling of Einstein's lift radiate or not?\\
      A: Well... viewed by an observer in the lift - NO!, as it is at rest; but, on the other hand, viewed by an observer on the Earth - YES!, as it falls down with constant acceleration $\vec{g}$.}\\
    Here, the confusion in the answers is of course due to the fact that one tries to apply a \emph{purely local} physical law - the equivalence principle, to the very nonlocal phenomenon, which is radiation in electromagnetic theory.

    This gives a hint as to how to define what is radiation in General Relativity. One can not expect that in this nonlinear theory radiation can be defined in terms of local notions. This point is raised, and consequently developed by {\bf Andrzej Trautman} in two papers \cite{T1,T2} submitted to \emph{Bulletin de l'Academie Polonaise des Sciences}, behind the Iron Curtain\footnote{Although the papers were published behind the Iron Curtain, their results were exposed to the western audience. In the next two months (May-June, 1958) after the submission of \cite{T1,T2} Trautman gave a series of lectures at King's College London presenting the theses of \cite{T1,T2}. The audience of his lectures included H. Bondi and F. Pirani, and the lectures were mimeographed \cite{kings} and spread among western relativists.}, in April 1958. This led him to finally solve our problems (4)-(5) in \cite{T2} and (6)-(7) in \cite{RT},  thereby switching the red light to green. 

\subsection{Green light: Trautman}
 Trautman's general idea in defining what a gravitational wave is in the full Einstein theory, was to say that it should satisfy certain boundary conditions at infinity. More precisely, from all spacetimes, i.e., solutions of Einstein's equations in the full theory, he proposed to select only those that satisfy boundary conditions at infinity, which were his \emph{generalizations} of Sommerfeld's radiation conditions \cite{CH}. These are known in the linear theory of a scalar field, and Trautman in \cite{T1,T2} generalizes them to a number of \emph{physical theories}. In paper \cite{T1} Trautman reformulates Sommerfeld's radiation boundary conditions for the scalar inhomogeneus wave equation into a form that is then generalized to other field theories. As an example he shows how this generalization works in Maxwell's theory, and that it indeed selects the outgoing radiative Maxwell fields from all solutions of Maxwell's equations.
 
 In the next paper \cite{T2} Trautman does the same for Einstein's General Relativity. In it Trautman defines the \emph{boundary conditions to be imposed on gravitational fields due to isolated systems of matter} (\cite{T2}, on p. 409, equations (9) and (10), see also the Appendix below). This is the first step in solving our problems (4) and (5).

 He then passes to the treatment of our problem (5). He uses the \emph{von Freud superpotential} 2-form $\mathcal F$ \cite{vF}, to split the Einstein tensor $E$ into $E=\der{\mathcal F}-\kappa \mathfrak{T}$,  so that the Einstein equations $E=\kappa T$ take the
   form  $$\der {\mathcal F}=\kappa (T+\mathfrak{T}).$$
   Here $T$ is the energy-momentum 3-form (see the Appendix for more details), and $\kappa$ is a constant related to the gravitational constant $G$ and the speed of light $c$ via $\kappa=\tfrac{8\pi G}{c^4}$ (in the following we work with physical units in which $c=1$).

   Since $\mathfrak{T}$ is a 3-form totally determined by the geometry, it is interpreted as the energy-momentum 3-form of \emph{pure gravity} (\cite{T2}, on p. 407, equations (1) and (2)). The closed 3-form $T+\mathfrak{T}$ is then used to define the 4-momentum $P^{\mu}(\sigma)$ of a \emph{gravitational field attributed to every space-like hypersurface} $\sigma$ of a spacetime satisfying his radiative boundary conditions, (\cite{T2}, p. 408, equation (5)). He shows that $P^\mu(\sigma)$ is \emph{finite} and \emph{well defined}, i.e., that it does not depend on the coordinate systems adapted to the chosen boundary conditions, (\cite{T2}, pp. 409-410). Using his boundary conditions he then calculates how much of the gravitational energy $p^{\mu}= P^{\mu}(\sigma_1)-P^{\mu}(\sigma_2)$ contained between the spacelike hypersurfaces $\sigma_1$ (initial one) and $\sigma_2$ (final one) \emph{escapes to infinity} (\cite{T2}, p.410-411, equations (16)-(17)).

   Finally, he shows that $p^0$ is \emph{nonnegative}, (\cite{T2}, p. 411, remark after (17)), saying that radiation is present when $p^0>0$.  

   Taken together, everything we have said so far about Trautman's results from \cite{T2}, \emph{solves our problems} (4) and (5): What in popular terms is called a \emph{gravitational wave in the full GR theory} is a \emph{spacetime satisfying Trautman's boundary conditions with $p^0>0$}; the \emph{energy of a gravitational wave} contained between hypersurfaces $\sigma_1$ and $\sigma_2$ is given by $p^0$.

   Trautman in \cite{T2} proves only that $p^0\geq 0$. If the inequality was sharp, $p^0>0$, this would give a proof of the statement that spacetimes satisfying Trautman's boundary conditions, or better said, the gravitational waves associated with them, \emph{carry energy}. Trautman in \cite{T2} did not have such a proof. To handle this problem, one can try to find an example of an \emph{exact solution} to the Einstein equations satisfying Trautman's boundary conditions, and to show that in this example $p^0$ is \emph{strictly} greater than zero. This approach is taken by I. Robinson and Trautman in \cite{RT}, and we will comment on this later.

As regards the paper \cite{T2}, it is worthwhile to mention also, that Trautman in addition shows there two other interesting things implied by his boundary conditions. The first of them  is the fact that in the presence of electromagnetic radiation a spacetime satisfying his boundary condtions has Ricci tensor, which far from the sources, is in the form of a \emph{null dust} $R_{\mu\nu} = \rho k_\mu k_\nu$, with $k$ being a \emph{null vector}, (\cite{T2}, p. 411, eq. (20)). This in particular means that the electromagnetic/gravitational radiation in his spacetimes travels with the speed of light. The second interesting feature he shows is that \emph{far from the sources the Riemann tensor of a spacetime satisfying his radiative boundary conditions is of Petrov type} $N$, (\cite{T2}, p. 411, eq. (21)).
Since far from the sources $Riemann=Weyl$, this verifies the \emph{intuition} of Pirani from \cite{Pi}: spacetimes satisfying radiative boundary conditions satisfy the algebraic speciality criterion, and from all the possibilities of algebraic speciality they choose a type $N$ Weyl tensor as the leading term at infinity. This was later developed into the celebrated \emph{peeling-off theorem} attributed to {\bf Ray Sachs} \cite{Sa}. 

The last two of our problems (6)-(7) were adressed by I. Robinson and Trautman in \cite{RT}. There they \emph{found a large class of exact solutions} of the full system of Einstein equations satisfying Trautman's boundary conditions. The solutions describe waves with \emph{closed fronts} so \emph{they can be interpreted as coming from bounded sources}. Explicitly, the Robinson-Trautman waves are given by the following formulae:
\be\begin{aligned}
g&=\frac{2r^2\der\zeta\der\bar{\zeta}}{P^2}-2\der u \der r-\Big(\triangle\log P-2r(\log P)_u-\frac{2m}{r}\Big)\der u^2\\
&\triangle\triangle(\log P)+12m(\log P)_u-4m_u=0,\quad\triangle=2P^2\partial_\zeta\partial_{\bar{\zeta}},\\
&\quad\quad\quad\quad\quad P=P(u,\zeta,\bar{\zeta}),\quad m=m(u).\end{aligned}\label{rt1}\ee
\emph{These solutions solve our last two problems} (6) \emph{and} (7). For some of them $p^0>0$, so they correspond to gravitational waves that \emph{do carry energy} (see \cite{Ch1,Ch2,Ch3}).

Concluding this section we say that the Bondi-Pirani-Robinson papers \cite{B,BPR} and the Trautman-Robinson papers \cite{T2,RT} solve all our problems (1)-(7), giving the green light to further research on gavitational radiation. We will not comment on these further developments since they are well documented, see e.g. \cite{K}.

\section{APPENDIX: DETAILS OF TRAUTMAN's FUNDAMENTAL PAPER}
Since Trautman's original paper \cite{T2} is published in a journal that is not very easily accessible we provide here a brief self-contained exposition of its content. 
\subsection{Trautman's radiative boundary conditions \cite{T1,T2}}
Let $T$ be a world tube containing all matter in a spacetime $M$. We assume that there exists a larger world tube $T_R$ containing tube $T$, $T\varsubsetneq T_R$ with a boundary cylinder $\Sigma_R$, such that $M_R=M\setminus T_R$ has the following properties: (a) $M$ is foliated\footnote{Here, for clarity of presentation, we assume the existence of a foliation by spacelike hypersurfaces. Trautman in \cite{T2} does not need it. For his purposes only two spacelike hypersurfaces, the initial one $\sigma_1$ and the final one $\sigma_2$, are neccessary. Again, for clarity of presentation, we introduced the tube $T_R$ which is not present in Trautman's paper.} by spacelike hypersurfaces $\sigma$, each of which has the topology of $\bbR^3$, and (b) on each $\sigma$ every geodesic tangent to $\sigma$ (in the induced metric on $\sigma$), which starts at $\Sigma_R$ and is tangent to the outward normal to $\Sigma_R$, escapes to infinity on $\sigma$.

Let $p$ be a point in $M_R$. It belongs to one of the  hypersurfaces $\sigma$. We define a function $r$ on $M_R$ assigning to $p$ its geodesic distance, along $\sigma$ and in the induced metric on $\sigma$, to the sphere $S_R=\Sigma_R\cap\sigma$. For $r>R$ we then have a foliation of $M$ by the cylindrical boundaries $\Sigma_r$ of the world tubes $T_r$, $r>R$, whose intersections with each $\sigma$ is a sphere $S_r$.

Since we will be interested in the asymptotics as $r\to \infty$, we will be only concerned with radii $r$ much larger than $R$. In particular we will be interested in integral quantities, such as fluxes, defined at $r=\infty$. For this we will need the spheres $S_r$ and the cylinder $\Sigma_r$ as $r\to \infty$. Let us also remark that we allow the spacelike hypersurfaces $\sigma$ to be \emph{asymptotically null} as $r\to\infty$.

Let $n$ be the tangent vector to each $\sigma$ which is the outward unit normal to $S_r$ , and $t$ be the future pointing unit normal to each $\sigma$. Define $k$ to be  $k=t+n$, so that $k$ is a future pointing \emph{null} vector (we assume that the speed of light $c=1$). 

A spacetime $M$ satisfies \emph{radiative boundary conditions} \cite{T2} if there exist $r_0>R$ and a coordinate system $(x^\mu)$ in $M\setminus T_{r_0}$ in which the metric $g_{\mu\nu}$ and its derivatives $g_{\mu\nu,\rho}$ satisfy the following asymptotics as $r\to\infty$:   
\be \begin{aligned}
  g_{\mu\nu}&=\eta_{\mu\nu}+O(r^{-1}),\quad\quad g_{\mu\nu,\rho}=h_{\mu\nu}k_\rho+O(r^{-2}),\\
  &\Big(h_{\mu\nu}-\tfrac12\mathrm{Tr}(h)\eta_{\mu\nu}\Big)k^\nu=O(r^{-2}).\end{aligned}\label{bc}\ee
Here $\eta_{\mu\nu}$ is the Minkowski metric $\eta_{\mu\nu}=\mathrm{diag}(1,-1,-1,-1)$, $h_{\mu\nu}$ are functions on $M\setminus T_{r_0}$ with the asymptotics 
$$h_{\mu\nu}=O(r^{-1}),$$
and the symbol $\mathrm{Tr}(h)=\eta^{\mu\nu}h_{\mu\nu}$.

We note that among the equations (\ref{bc}), which provide the definition, the equations in the first line are generalizations of Sommerfeld's radiation condition, and the equations in the second line guarantee that the postulated coordinate system is asymptotically harmonic. It is this last condition that is crucial for the energy momentum tensor of a pure gravitational field $\mathfrak{T}^\mu$ (defined in the next section) to be of the form (\ref{fo1}), which guarantees radiation (see below).  
\\

\subsection{Definition of Trautman's energy}
We rewrite the Einstein equations
\be R_{\mu\nu}-\tfrac12 Rg_{\mu\nu}=\kappa T_{\mu\nu}\label{eine}\ee in terms of differential forms\footnote{This is not done in Trautman's paper. He uses tensorial notation. Although the formulation in terms of tensor-valued differential forms is not in \cite{T2}, the presentation here is equivalent to the one from the original paper \cite{T2}. It is also due to Trautman. He was using it already in the 1970's.}. For this we use a coframe $\theta^\mu$ in spacetime, in which the spacetime metric is $g=g_{\mu\nu}\theta^\mu\theta^\nu$. Using it we define the Levi-Civita connection 1-form $\Gamma^\mu{}_\nu$, which satisfies
    $$\begin{aligned}
      \der\theta^\mu+\Gamma^\mu{}_\nu\dz\theta^\nu=0,\\
      \der g_{\mu\nu}-\Gamma^\rho{}_\mu g_{\rho\nu}-\Gamma^\rho{}_\nu g_{\mu\rho}=0,\end{aligned}$$
      the curvature 2-form
      $$\Omega^\mu{}_\nu=\der\Gamma^\mu{}_\nu+\Gamma^\mu{}_\rho\dz\Gamma^\rho{}_\nu=\tfrac12 R^\mu{}_{\nu\rho\sigma}\theta^\rho\dz\theta^\sigma,$$
 and a 3-form
      $$*\theta_\mu=\tfrac16\eta_{\mu\nu\rho\sigma}\theta^\nu\dz\theta^\rho\dz\theta^\sigma.$$
 Here $$\eta_{\mu\nu\rho\sigma}=\sqrt{|\mathrm{det}g_{\alpha\beta}|}~\epsilon_{\mu\nu\rho\sigma},$$ 
with $\epsilon_{\mu\nu\rho\sigma}$ being the totally skew symmetric symbol in 4-dimensions, satisfying $\epsilon_{0123}=1$. For further use we also need a 1-form $$\eta_{\mu\nu\rho}=\eta_{\mu\nu\rho\sigma}\theta^\sigma.$$

With this notation the Ricci tensor $R_{\mu\nu}$ and its scalar $R$ are
$$R_{\mu\nu}=R^\rho{}_{\mu\rho\nu},\quad\quad R=g^{\mu\nu}R_{\mu\nu},$$
and Einstein's equations (\ref{eine}) can be written in terms of 3-forms:
$$G^\mu=(R^{\mu\nu}-\tfrac12g^{\mu\nu}R)~*\theta_\nu\quad\mathrm{and}\quad T^\mu=T^{\mu\nu}*\theta_\nu.$$
We thus have that the Einstein equations (\ref{eine}) are equivalent to 
$$G^\mu=\kappa T^\mu.$$
It follows that the left hand side of these equations is linear in terms of the curvature 2-form $\Omega^\mu{}_\nu$. Explicitly, one shows that the Einstein equations are equivalent to
$$-\tfrac12\eta^\mu{}_\nu{}^\rho\dz\Omega^\nu{}_\rho=\kappa T^\mu.$$
It is this form of the Einstein equations that is suitable for defining the energy of gravitational field. Indeed, introducing the von Freud \cite{vF} superpotential 2-form ${\mathcal F}^\mu$ as
$${\mathcal F}^\mu=\tfrac12\eta^\mu{}_\nu{}^\rho\dz\Gamma^\nu{}_\rho,$$ and the 3-form
$$\mathfrak{T}^\mu=\tfrac{1}{2\kappa}(\der\eta^\mu{}_\nu{}^\rho\dz\Gamma^\nu{}_\rho+\eta^\mu{}_\nu{}^\rho\dz\Gamma^\nu{}_\sigma\dz\Gamma^\sigma{}_\rho),$$
we see that the Einstein equations become
\be \der{\mathcal F}^\mu=\kappa (T^\mu+\mathfrak{T}^\mu).\label{eine1}\ee
They imply a conservation law
$$\der(T^\mu+\mathfrak{T}^\mu)=0.$$
This enables to interpret the quantity $T^\mu+\mathfrak{T}^\mu$ as the energy momentum of both the matter ($T^\mu$) and the pure gravitational field ($\mathfrak{T}^\mu$). Note that $\mathfrak{T}^\mu$ is expressible in terms of the components of geometry, such as the coframe $\theta^\mu$, metric $g$ and the Levi-Civita connection $\Gamma^\mu{}_\nu$, but not in terms of matter.

We are now in a position to define the total 4-momentum $P^\mu [\sigma]$ assigned to a given spacelike hupersurface $\sigma$ \cite{T2}. This is

$$P^\mu[\sigma]=\int_{\sigma} (T^\mu+\mathfrak{T}^\mu)=\tfrac{1}{\kappa}\lim_{r\to\infty}\oint_{S_r}{\mathcal F}^\mu,$$
where we have used Stokes' theorem, the Einstein equations (\ref{eine1}), and the notation of the previous section.

Staying with this notation we now return to the world tube $T_r$ in a \emph{spacetime satisfying radiative boundary conditions}. We assume that $r>r_0$, and consider two spacelike hypersurfaces $\sigma_1$ and $\sigma_2$ from the foliation considered in the previous section. We further assume that $\sigma_1$ is earlier than $\sigma_2$ and consider the 4-dimensional region of spacetime contained in the `can' $T_r(\sigma_1,\sigma_2)$ with bottom $\sigma_1$, top $\sigma_2$, and lateral part $\Sigma_r(\sigma_1,\sigma_2)$ being the part of $\Sigma_r$ cut out by $\sigma_1$ and $\sigma_2$. Using Stokes' theorem again we have
$$\begin{aligned}
  0=&\int_{T_r(\sigma_1,\sigma_2)}\der(T^\mu+\mathfrak{T}^\mu)=\lim_{r\to \infty}\oint_{\partial(T_r(\sigma_1,\sigma_2))}(T^\mu+\mathfrak{T}^\mu)\\
  &=-P^\mu[\sigma_1]+P^\mu[\sigma_2]+\lim_{r\to\infty}\int_{\Sigma_r(\sigma_1,\sigma_2)}(T^\mu+\mathfrak{T}^\mu).\end{aligned}$$
Denoting
$$p^\mu=\lim_{r\to\infty}\int_{\Sigma_r(\sigma_1,\sigma_2)}(T^\mu+\mathfrak{T}^\mu)=\lim_{r\to\infty}\int_{\Sigma_r(\sigma_1,\sigma_2)}\mathfrak{T}^\mu,$$
with the last equality being implied by the fact that when $r\to\infty$ the matter energy momentum $T^\mu=0$, we get
$$p^\mu=P^\mu[\sigma_1]-P^\mu[\sigma_2].$$
Obviously, $p^\mu$ is interpreted as the \emph{4-momentum radiated from the spacetime to spatial or null infinity between the hypersurfaces} $\sigma_1$ and $\sigma_2$ \cite{T2}.  For radiation we need $p^\mu\neq 0$, and for radiation \emph{escaping} to infinity we need $p^0>0$. We call $p^0$ \emph{Trautman's energy loss due to radiation}.

\subsection{Trautman's green light theorems}
    Trautman did not collect the results of his papers \cite{RT,T2} into displayed theorems. We do it here for the convenience of the reader.\\

\noindent    
\emph{Theorem 1} \cite{T2}\\
    Assume that the spacetime satisfies the radiative boundary conditions (\ref{bc}). Then
    \begin{itemize}
    \item its purely gravitational energy momentum $\mathfrak{T}^\mu$ is 
      \be \mathfrak{T}^\mu=\tau k^\mu k_\nu *\der x^\nu+O(r^{-3}),\label{fo1}\ee
      with the function $\tau$ \emph{nonnegative},
    \item Trautman's energy loss due to radiation $p^0$ does not depend on the choice of coordinate system from the class adapted to the boundary conditions (\ref{bc}),
\item it depends only on the choice of the spacelike hypersurfaces $\sigma_1$ and $\sigma_2$,
\item $p^0$ is \emph{nonnegative}, so that there is radiation in the system, if $p^0\neq 0$,
\item this is true even if the hypersurfaces $\sigma_1$ and $\sigma_2$ are \emph{asymptotically null} as $r\to\infty$,
          \item if, in addition, an electromagnetic field is present the Ricci tensor is of the form $$R_{\mu\nu}=\rho k_\mu k_\nu+O(r^{-2}),\quad \rho=O(r^{-2}),$$
      \item the Riemann tensor of the spacetime in the regions, where $r\to\infty$ is of type~$N$, $$k_{[\mu} R_{\nu\rho]\sigma\tau}\to 0,\quad k^\mu R_{\mu\nu\rho\sigma}\to 0.$$
    \end{itemize}~\\

\noindent
\emph{Theorem 2} (joint work with I. Robinson \cite{RT})\\
There exists exact, explicit solutions of the vacuum Einstein equations, satisfying radiative boundary conditions. These are vacuum Robinson-Trautman solutions (\ref{rt1}) for which $[P^2(\log P)_{u\bar{\zeta}}]_{\bar{\zeta}}\neq 0$. For them $p^\mu\neq 0$ and $p^0>0$, i.e. they exhibit energy loss due to radiation.

\end{document}